\documentclass[letterpaper]{article}
\usepackage[T1]{fontenc}
\usepackage{geometry}
\usepackage{setspace}
\usepackage{authblk}
\usepackage{graphicx}

\usepackage[
backend=biber,
style=chem-acs,
sorting=none,
articletitle=true
]{biblatex}
\usepackage{amsmath}
\usepackage{gensymb}

\newcommand{\VOC}{V_{\text{OC}}}
\newcommand{\Vapp}{V_{\text{app}}}
\newcommand{\VMPP}{V_{\text{MPP}}}
\newcommand{\JSC}{J_{\text{SC}}}
\providecommand{\supercite}[1]{\textsuperscript{[\texttt{\detokenize{#1}}]}}
\newcommand{\Pmat}{\mathbf{P}}
\newcommand{\one}{\mathbf{1}}
\newcommand{\pol}{\mathrm{pol}}

\newcommand{\abs}{\mathrm{abs}}

\newcommand{\Pmix}{\mathbf{P}_{\mathrm{mix}}}
\newcommand{\mix}{\text{mix}}
\newcommand{\actst}{\text{active}}

\title{Emergent Non-Equilibrium Emission Profiles via\\Luminescent Angle Restriction for\\Photovoltaic Energy Conversion}
\author[1]{Jonathan McTague}
\author[2]{Rivi Ratnaweera}
\author[1,3]{Matthew Sheldon*}
\affil[1]{Department of Chemistry, University of California, Irvine, Irvine, California 92697, United States}
\affil[2]{Department of Chemistry, Texas A\&M University, College Station, Texas 77843, United States}
\affil[3]{Department of Materials Science and Engineering, University of California, Irvine, Irvine, California 92697, United States}

\date{*Email: m.sheldon@uci.edu}

\begin{document}

\maketitle

\begin{abstract}
The detailed-balance method determines photovoltaic conversion efficiencies by relating non-equilibrium converter operation to an equilibrium reference state. Conventionally, the angular and spectral radiative channels are assumed to be defined by the equilibrium absorptivity and emissivity, with the photon chemical potential setting their occupation. The spatial distribution of emission must also be considered when radiative energy is redistributed within a complex nanophotonic converter architecture. Here we analyze this redistribution with a proposed luminescent angle restrictor, in which a flat-plate photovoltaic absorber is coupled to an overlayer with vertically aligned nanorods. The nanorod layer is nearly transparent to near-normal sunlight, while oblique photovoltaic luminescence can be absorbed, re-emitted, and partially returned to the absorber. To calculate the operating emission profile, we develop a Markov-chain transport model with an absorbing-chain formulation that enforces microscopic reversibility for local optical transitions while determining the spatially distributed steady-state emission rates. The model recovers the flat-plate detailed-balance limit and predicts voltage-dependent external emission profiles from the coupled architecture. Under idealized conditions, this non-equilibrium angular redistribution increases conversion efficiency relative to the conventional flat-plate reference. We further evaluate nanorod photoluminescence quantum yield, angular restriction, orientational disorder, and Auger recombination to identify practical constraints on the proposed architecture. These results show how detailed-balance limits for photovoltaic structures can depend on internal radiative energy redistribution and the resulting non-equilibrium emission profile, beyond what can be inferred from the macroscopic equilibrium optical response alone.
\end{abstract}

\section*{Keywords}

photoluminescence, photon recycling, photovoltaic systems, detailed balance, absorbing Markov chains, semiconductor nanorods

\vspace{0.75\baselineskip}
\section{Introduction}

Improving the efficiency of solar energy converters remains one of the most cost-effective routes to large-scale deployment of solar power.\supercite{WoodhouseSunShot2016,SolarFuturesStudy2021} Since its publication in 1961, the detailed-balance method outlined by Shockley and Queisser\supercite{ShockleyQueisser} has served as the foundational framework for estimating photovoltaic (PV) efficiency limits. The power of the method comes from treating the cell as a converter that exchanges radiation with its environment, so that efficiency limits can be derived from macroscopic photon fluxes rather than from a microscopic description of the junction, doping, or carrier dynamics. Photons absorbed from the Sun generate electron-hole pairs (EHPs), radiative recombination emits photons back to the environment, and the imbalance between these fluxes can be extracted as electrical current. At thermal equilibrium, detailed balance and microscopic reversibility require every elementary process to be balanced by its reverse. For a single-junction solar cell, this condition links optical absorption and emission, or equivalently EHP generation and recombination. The key step in the Shockley-Queisser construction is an assumption about which aspects of this description of radiative exchange in equilibrium can be extended to an illuminated device driven out of equilibrium by sunlight.

For a cell in true thermal equilibrium with its surroundings, Kirchhoff's law requires the angular and spectral emissivity of the cell to match its absorptivity, so that there is no net radiative exchange with the surrounding blackbody field.\supercite{Kats} Under solar illumination, an additional generation flux of EHPs is introduced, and the non-equilibrium carrier population is described by a quasi-Fermi level splitting, or photon chemical potential, $\mu_{\gamma}=q\Vapp$, where $\Vapp$ is the operating electrical voltage of the cell. In the standard detailed-balance treatment, this voltage is assumed to modify the occupation of the optical modes defined by the equilibrium response, rather than change which angular or spectral channels are available for emission. This assumption has clear physical intuition for a conventional flat-plate solar cell. The device optics determine which modes can absorb and emit, while the non-equilibrium carrier population primarily controls the occupation of those modes.\supercite{Rau2007}

The same logic also explains why luminescence is central to high photovoltaic performance. As emphasized by Green and by Miller and co-workers,\supercite{GreenRadiative,Kurtz} a high-efficiency solar cell should be an efficient light emitter at open circuit, because strong radiative emission indicates that photogenerated carriers are maintained at high quasi-Fermi level splitting rather than lost through nonradiative recombination. The angular distribution of that emission is also important. Restricting radiative escape to the small solid angle from which sunlight is received, while preserving solar absorption, can increase the operating voltage and extractable free energy.\supercite{RauKirchartz2014} This principle has motivated optical structures such as parabolic mirrors or apertures that restrict luminescent escape from solar cells while preserving solar absorption.\supercite{PolmanAtwater2012,RiviDB} For direct sunlight in the ideal radiative limit, angular restriction and geometric solar concentration can provide equivalent voltage enhancements by reducing the mismatch between the étendue of incident and emitted radiation. Concentrating optics broaden the angular range of sunlight at the cell, while angular restriction narrows the range into which luminescence can escape, establishing a correspondence that allows angular restriction to be expressed as an equivalent solar concentration factor.

More recently, the detailed-balance framework has been applied to more complex energy-conversion architectures with nanophotonic elements whose absorptivity can depend strongly on angle, frequency, polarization, or subwavelength field structure.\supercite{Garnett,Munday} These studies show how efficiency limits can change when the optical response of a solar cell is no longer that of a simple flat plate. In such treatments, the overall device is described by a fixed set of optical channels defined by its equilibrium absorptivity-emissivity relation. As in the flat-plate case, the photon chemical potential in the PV, $\mu_{\gamma}=q\Vapp$, changes the occupation of those radiative channels, rather than modifying which channels are available for emission. This description is powerful and well motivated when the optical structure acts passively, selecting absorption and escape modes. When nanoscale optical elements also promote radiative exchange, photon recycling, or re-emission within the architecture before light escapes the device, the spatial distribution of local emission rates must also be considered. In that case, local detailed balance still governs each elementary optical transition, while the external emission profile depends on both the fixed optical response and the spatially distributed emission within the architecture. As we show here, this emission distribution emerges from the coupled non-equilibrium steady state of the full device and can change with operating voltage. The external emission is then obtained by linearly combining these local contributions with their corresponding angular escape probabilities.

To make this possibility concrete, we propose and analyze a luminescent angle restrictor (LAR) geometry, in which a conventional flat-plate photovoltaic cell is coupled to a vertically aligned layer of luminescent semiconductor nanorods (Figure~\ref{fig:lar-architecture}). The nanorod layer is nearly transparent to direct, near-normal sunlight, but can absorb and re-emit p-polarized oblique luminescence from the photovoltaic layer. In this geometry, the angular distribution of light leaving the device results from radiative exchange between the PV and the nanorod layer before photons escape to the environment. We describe this coupled radiative exchange with a Markov-chain model employing an absorbing-chain formulation that determines the spatially distributed steady-state emission rates and their contributions to external emission while preserving microscopic reversibility for local optical transitions. The model resolves how the depth-dependent emission rates and external angular emission profile vary with operating voltage (Figure~\ref{fig:voltage-emission}). Our objective is to determine whether, and under what optical and material conditions, this coupled architecture can improve photovoltaic performance relative to a flat-plate reference under matched illumination and angular-restriction conditions.

We first benchmark the model against the conventional detailed-balance result in the flat-plate limit (Figure~\ref{fig:benchmark-bandgap}). We then examine how nanorod absorption strength, PV--LAR refractive-index contrast, and angular restriction affect photon recycling and conversion efficiency. Finally, we assess the constraints imposed by nonunity nanorod photoluminescence quantum yield (QY), imperfect nanorod alignment, and Auger recombination in the PV.

\section{Results and Discussion}

\begin{figure}
    \centering
    \includegraphics[width=1\linewidth]{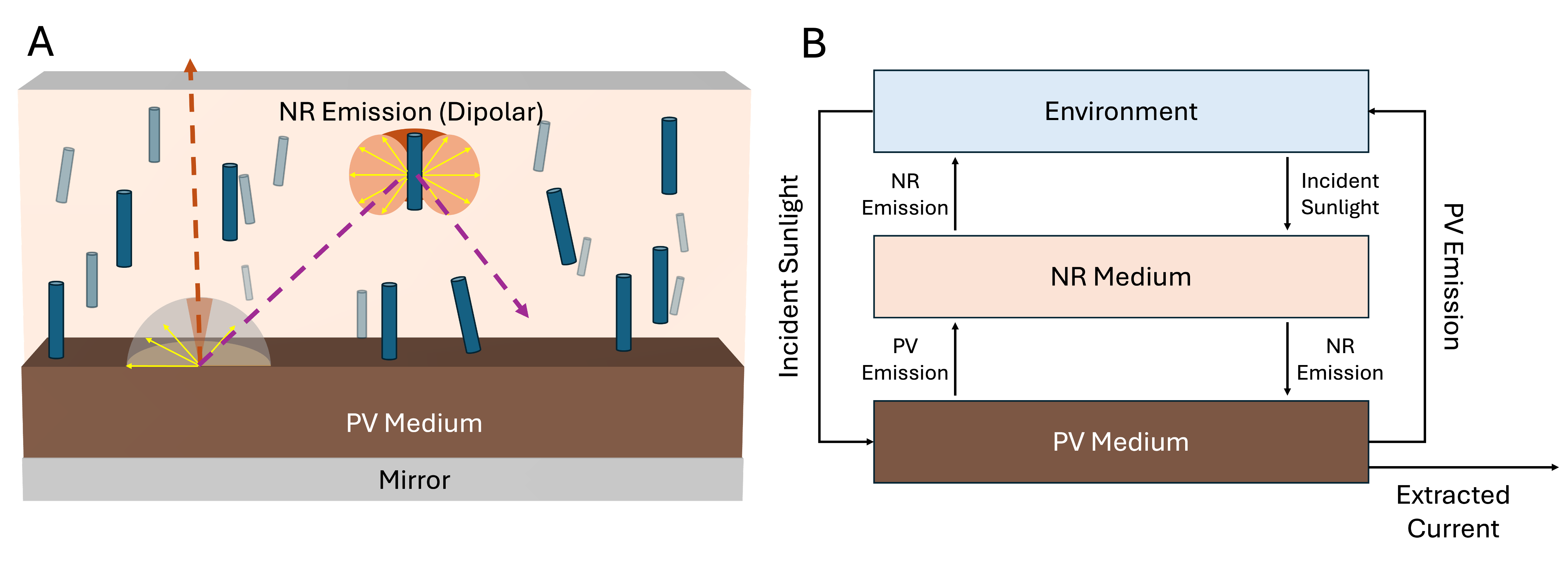}
    \caption{(A) Schematic of a luminescent angle restrictor (LAR) coupled to a conventional flat-plate photovoltaic cell. An ideal mirror is placed at the bottom of the coupled device. (B) Photon-transport pathways included in the model. Photons enter through absorption of incident sunlight and exit through radiative emission. Electron-hole pairs generated within the PV may alternatively be extracted as current instead of undergoing radiative recombination.}
    \label{fig:lar-architecture}
\end{figure}

Figure~\ref{fig:lar-architecture}A shows the idealized LAR module used in the calculations below. A flat-plate PV cell is coated with a transparent dielectric layer containing vertically aligned semiconductor nanorods (NRs), with a perfectly reflective mirror at the bottom of the PV to suppress transmission. The refractive indices of the PV and LAR medium are denoted by $n_P$ and $n_L$, respectively. Unless otherwise noted, the PV is treated as ideal and absorbs all incident photons with energy $\geq E_g$ with unity probability and no nonradiative loss. The embedded NRs are assigned the same band gap as the PV and unity photoluminescence quantum yield (QY). NRs are incorporated because they can approximate linearly polarized dipolar radiators along their long axis,\supercite{Alivisatos_lin_pol} giving a dipolar emission pattern proportional to $\sin^2\theta$, where $\theta$ is the angle within the LAR measured from the vertical NR axis. Similar anisotropic optical responses have been characterized experimentally in core/shell CdS/CdSe and CsPbBr$_3$ nanorod ensembles.\supercite{RiviAlign,RodriguezOrtiz2023} For ideally vertical NRs, absorption is polarization selective: $\alpha_{L,p}(\theta)=A_0\sin^2\theta$ and $\alpha_{L,s}(\theta)=0$. Here, $A_0$, in $\mathrm{m}^{-1}$, sets the nanorod absorption-coefficient scale; for a fixed absorption cross-section, increasing $A_0$ corresponds to increasing nanorod loading. Thus, the s-polarized half of unpolarized PV luminescence remains unabsorbed by the NRs on each pass, regardless of $A_0$. 

At first, it may seem counterintuitive to include an optical layer that can absorb sunlight before it reaches the PV. For direct sunlight incident near the surface normal, however, a vertically aligned NR layer absorbs weakly, so most incident light passes into the PV. As summarized in Figure~\ref{fig:lar-architecture}B, absorption in the PV generates electron-hole pairs that can contribute to extracted current or recombine radiatively, producing luminescence over a broad range of angles. The p-polarized component of oblique PV luminescence interacts more strongly with the NR layer, where it can be absorbed and re-emitted, including back toward PV-absorbing channels. The angular distribution of light leaving the device is therefore shaped by radiative exchange between the PV and NR layer before photons escape to the environment. In this sense, the LAR modifies the angular distribution of escaping luminescence in a manner similar to external angle-restricting optics such as parabolic mirrors or angle-dependent distributed Bragg reflectors,\supercite{Atwater,DBR_Xu,AtwaterParabolic,GaAs_Angle,PV_enhance,RiviDB} but the angular redistribution is generated internally by the coupled luminescent architecture.

This distinction is important because the external emission from the coupled device depends not only on its angle-dependent absorptivity $a(\theta,\phi,E)$ and the PV quasi-Fermi level splitting, but also on the spatial distribution of emission within the NR layer. The emitted light reflects a voltage-dependent mixture of direct PV emission and NR emission after absorption and re-emission events. Changing $\Vapp$ changes the magnitude of PV luminescence that drives the NR layer and can change the relative contributions of these channels along the power curve. We therefore calculate the operating emission profile from the coupled radiative transport pathways, with the optical properties held fixed and the local emission contributions combined linearly using their corresponding angular escape probabilities. The resulting emission profiles are summarized in Figure~\ref{fig:voltage-emission} for the case in which the design is illuminated with direct sunlight at normal incidence.

\begin{figure}
    \centering
    \includegraphics[width=1\linewidth]{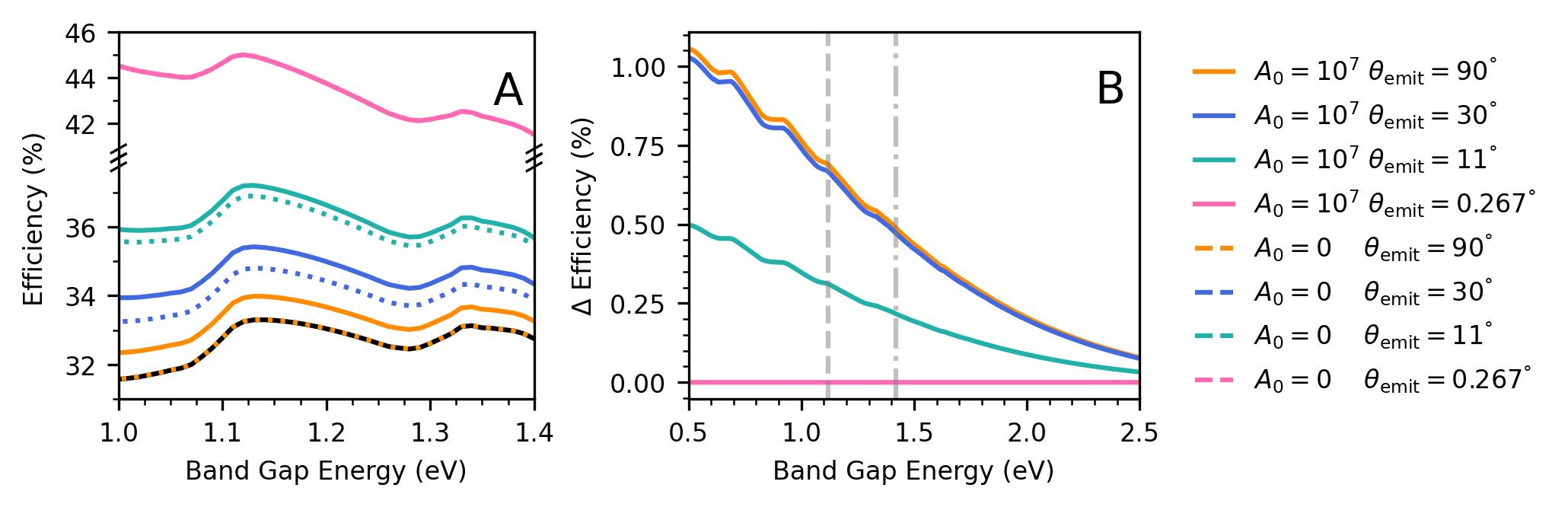}
    \caption{Benchmark and idealized LAR performance as a function of PV band-gap energy, $E_g$. Panel (A) compares conversion efficiency for a conventional flat-plate solar cell calculated using detailed balance (black) and the Markov-chain model (orange, dotted). The overlap of these traces provides the flat-plate benchmark. The solid angle of emission is restricted to a cone spanning polar angles $0 \leq \theta \leq \theta_{\mathrm{emit}}$ measured from the surface normal. Additional traces correspond to limiting emission half-angles $\theta_{\mathrm{emit}} = 90^\circ$ (orange), $30^\circ$ (blue), $11^\circ$ (turquoise), and $0.267^\circ$ (pink), equivalent to solar concentration factors of 1x, 4x, 27x, and 46,000x, respectively. Solid and dotted traces correspond to absorption coefficients of the LAR medium, $A_0 = 10^7\,\mathrm{m}^{-1}$ and $A_0 = 0\,\mathrm{m}^{-1}$, respectively. Panel (B) provides an expanded view over $E_g = 0.5$--$2.5\,\mathrm{eV}$ and indicates the absolute differences in conversion efficiency, in percentage points, between these two configurations, using the same color convention for $\theta_{\mathrm{emit}}$. Vertical dotted and dash-dotted lines mark the band gaps of Si, $E_g = 1.12\,\mathrm{eV}$, and GaAs, $E_g = 1.42\,\mathrm{eV}$, respectively. Calculations use the AM1.5D solar spectrum and assume a LAR refractive index $n_L = 1$ and PV refractive index $n_P = 1$.\supercite{NREL}}
    \label{fig:benchmark-bandgap}
\end{figure}

We benchmarked the Markov-chain calculation against the conventional flat-plate detailed-balance result. In this reference case, the nanorod layer is made optically inactive by setting its absorption coefficient $A_0 = 0~\mathrm{m}^{-1}$, with $n_P=n_L=1$ and unrestricted emission into the upper hemisphere ($\theta_{\mathrm{emit}}=90^\circ$). The agreement between the black detailed-balance curve and the orange dotted Markov-chain curve in Figure~\ref{fig:benchmark-bandgap}A confirms that the model recovers the standard flat-plate limit before nanorod coupling is introduced.

When the LAR layer is populated with nanorods, the calculation predicts that conversion efficiency can increase above the corresponding inactive-layer reference. In Figure~\ref{fig:benchmark-bandgap}A, the solid curves ($A_0=10^7~\mathrm{m}^{-1}$) are compared with the dotted curves ($A_0=0$) at each emission-cone half-angle, with $n_P=n_L=1$ throughout. Stronger angular restriction raises the total conversion efficiency of both configurations. For this index-matched geometry, the imposed angular restriction corresponds to an equivalent solar concentration factor $C_{\mathrm{eq}}=1/\sin^2\theta_{\mathrm{emit}}$. The smallest half-angle considered, $0.267^\circ$, approximately matches the Sun's angular radius and corresponds to $46{,}000\times$, the ideal maximum for three-dimensional geometric concentration of direct sunlight in air.

Figure~\ref{fig:benchmark-bandgap}B isolates the additional benefit of the LAR by plotting the efficiency gain, in percentage points, relative to the inactive-LAR reference at the same emission-cone half-angle. Over the band-gap range \(E_g=0.5\)--\(2.5~\mathrm{eV}\), the gain is greatest at lower band gaps and with weaker angular restriction, corresponding to lower equivalent solar concentration factors. The gain decreases with increasing band gap and stronger angular restriction. At \(\theta_{\mathrm{emit}}=0.267^\circ\), the additional benefit is negligible in these idealized calculations. Here, angular restriction already approximately matches the permitted emission étendue to that of the incident sunlight. The vertical markers identify the band gaps of Si and GaAs, providing reference points for these trends.

\begin{figure}
    \centering
    \includegraphics[width=1\linewidth]{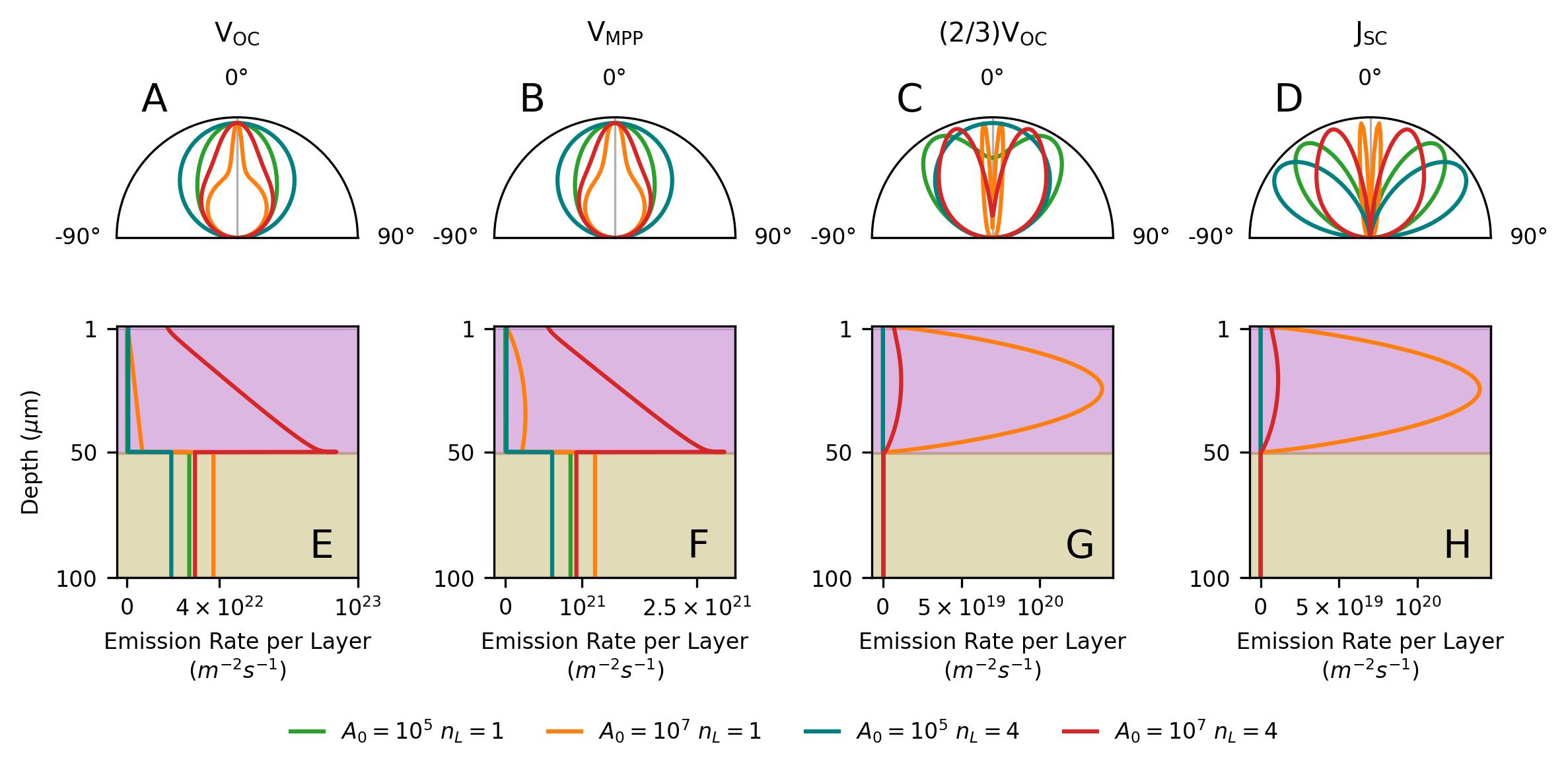}
    \caption{Voltage-dependent emission from a flat-plate PV coupled to a LAR medium. (A-D) External angular emission profiles at $\text{V}_{OC}$, $\text{V}_{MPP}$, (2/3)$\text{V}_{OC}$, and $\text{J}_{SC}$, respectively, for four configurations: $\text{A}_0 =10^5$ $\text{m}^{-1}$ and $n_L$ = 1 (green), $\text{A}_0 =10^7$ $\text{m}^{-1}$ and $n_L$ = 1 (orange), $\text{A}_0 =10^5$ $\text{m}^{-1}$ and $n_L$ = 4 (turquoise), and $\text{A}_0 =10^7$ $\text{m}^{-1}$ and $n_L$ = 4 (red). (E-H) Corresponding depth-dependent radiative emission rate profiles. The NR and PV media are shaded purple and tan, respectively. Calculations use direct solar illumination without solar concentration at normal incidence, $E_g$ = 1.12 eV, ideal vertical nanorod alignment, and unity photoluminescence quantum yield (QY).}
    \label{fig:voltage-emission}
\end{figure}

\begin{figure}
    \centering
    \includegraphics[width=1\linewidth]{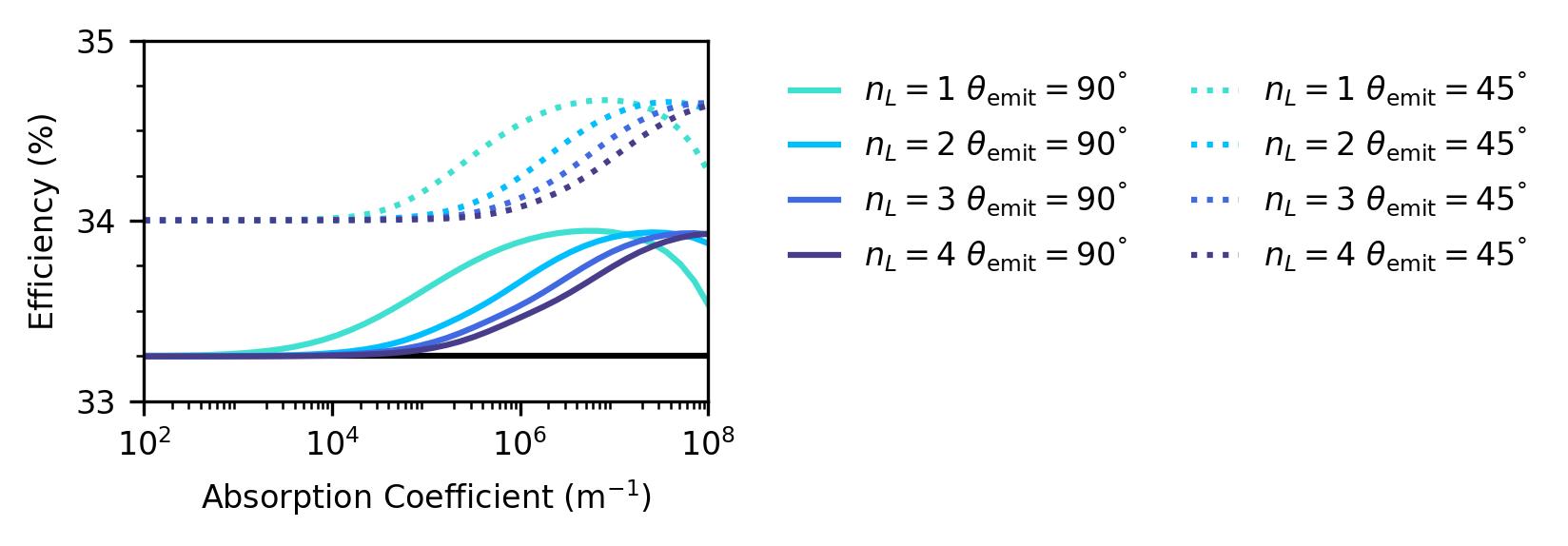}
    \caption{LAR performance as a function of nanorod loading and refractive index. Conversion efficiency plotted versus the absorption coefficient of the LAR medium, $A_0$, for LAR refractive indices $n_L = 1$ (cyan), $n_L = 2$ (light blue), $n_L = 3$ (blue), and $n_L = 4$ (dark blue). The solid angle of emission is restricted to a cone spanning polar angles $0 \leq \theta \leq \theta_{\mathrm{emit}}$ measured from the surface normal. Solid and dotted traces correspond to limiting emission half-angles $\theta_{\mathrm{emit}} = 90^\circ$ (1x solar concentration factor) and $\theta_{\mathrm{emit}} = 45^\circ$ (2x solar concentration factor), respectively. The flat-plate detailed-balance reference is provided (black).}
    \label{fig:ideal-design-space}
\end{figure}

Figure~\ref{fig:voltage-emission} provides the central evidence that the LAR-coupled device changes its angular emission profile with operating voltage, rather than varying only in overall emission intensity. These plots compare the external angular emission profiles (A--D) and depth-dependent radiative emission rates (E--H) at four operating points: open circuit ($\VOC$), maximum power point ($\VMPP$), $(2/3)\VOC$, and short circuit ($\JSC$). The four configurations combine $A_0=10^5$ and $10^7~\mathrm{m}^{-1}$ with $n_L=1$ and $4$. Refractive-index contrast changes the angular channels available for photon transport across the PV--LAR and LAR--environment interfaces, thereby modifying the balance between interlayer photon exchange, recycling, and escape. We examine the resulting dependence of conversion efficiency on refractive index systematically in Figure~\ref{fig:ideal-design-space}. The model includes multiple absorption, re-emission, reflection, and escape events, while the external emission is obtained by linearly combining the local steady-state emission rates with their corresponding angular escape probabilities. The optical properties are held fixed as the operating voltage changes. See the Methods section for full calculation details.

Nanorod concentration and refractive index shape this voltage-dependent angular response. At $\VOC$ and $\VMPP$, the external emission profiles are strongest near the surface normal. As $\Vapp$ decreases, PV radiative emission is suppressed and the external profiles develop off-axis lobes associated with nanorod emission. At short circuit, higher nanorod concentration shifts the characteristic lobes closer to the surface normal for both refractive indices. Lower-concentration LARs exhibit broader angular profiles, with the broadest short-circuit profile occurring for $n_L=4$.

Panels E--H of Figure~\ref{fig:voltage-emission} show the corresponding radiative emission rate per layer, defined as the number of photons emitted from each discretized layer per unit device area per unit time, in photons $\mathrm{m}^{-2}\,\mathrm{s}^{-1}$. At both $\VOC$ and $\VMPP$, the high-concentration, high-index configuration ($A_0=10^7~\mathrm{m}^{-1}$, $n_L=4$) has the largest emission rates within the LAR, with the strongest emission near the PV interface. At the lower-voltage points, the low-index, high-concentration configuration ($n_L=1$) has the largest LAR emission rates, with a maximum within the NR layer. The different horizontal scales show that these spatial changes accompany a substantial decrease in absolute emission rates as the voltage falls. These spatially varying rates determine the weights of the local contributions to external emission. Their implications for conversion efficiency depend on the balance between photon return to the PV, electrical extraction, and escape.

Among the four configurations in Figure~\ref{fig:voltage-emission}, the high-loading, low-index LAR ($A_0=10^7~\mathrm{m}^{-1}$, $n_L=1$) produces the emission most strongly concentrated near the surface normal at both $\VOC$ and $\VMPP$. This directionality is consistent with the larger efficiency benefit calculated for this configuration. The physical connection is analogous to conventional angle-restricting optics: suppressing radiative escape at a given operating voltage, while preserving solar absorption, supports a higher voltage and greater extractable electrical power. In the LAR, this directional response emerges from anisotropic absorption and re-emission coupled with interfacial photon transport and recycling.

Figure~\ref{fig:ideal-design-space} isolates the ideal design dependence on nanorod loading and refractive index. Conversion efficiency is plotted versus $A_0$ for $n_L=1$--$4$, with solid and dotted curves corresponding to emission-cone half-angles of $90^\circ$ and $45^\circ$, respectively (1x and 2x equivalent solar concentration). In the low-loading limit near $A_0=10^2~\mathrm{m}^{-1}$ and below, the curves for all $n_L$ converge to the flat-plate efficiency at the corresponding emission-cone half-angle. Thus, the low-loading limit of the $45^\circ$ curves represents the angle-restricted flat plate without an active LAR, and their subsequent rise demonstrates an additional benefit beyond angular restriction alone.

Along the rising portions of the curves, increasing $A_0$ improves efficiency because stronger nanorod absorption captures more of the p-polarized component of oblique PV luminescence and promotes photon recycling. Lower-index LARs develop appreciable gains at smaller $A_0$ and generally give higher efficiency through the intermediate-loading range, while increasing $n_L$ shifts this response toward higher loading. This trend does not continue indefinitely: the $n_L=1$ curves reach a maximum and then decline, while higher-index configurations reach comparable efficiencies at larger $A_0$. The resulting crossings show that the preferred refractive index depends on loading. Within each angular-restriction family, the highest efficiencies shown are similar, with refractive index primarily changing the loading required to approach them.

The design rule is therefore to choose nanorod loading and refractive index together to enhance useful photon recycling while preserving solar absorption. Refractive-index contrast changes the angular channels available for photon transport across the PV--LAR and LAR--environment interfaces, modifying the balance between interlayer photon exchange, recycling, and escape. These coupled processes determine the loading range over which the LAR improves conversion efficiency. Although angular restriction raises the total efficiency, the additional LAR benefit is measured relative to the flat-plate reference with the same emission cone.

\begin{figure}
    \centering
    \includegraphics[width=1\linewidth]{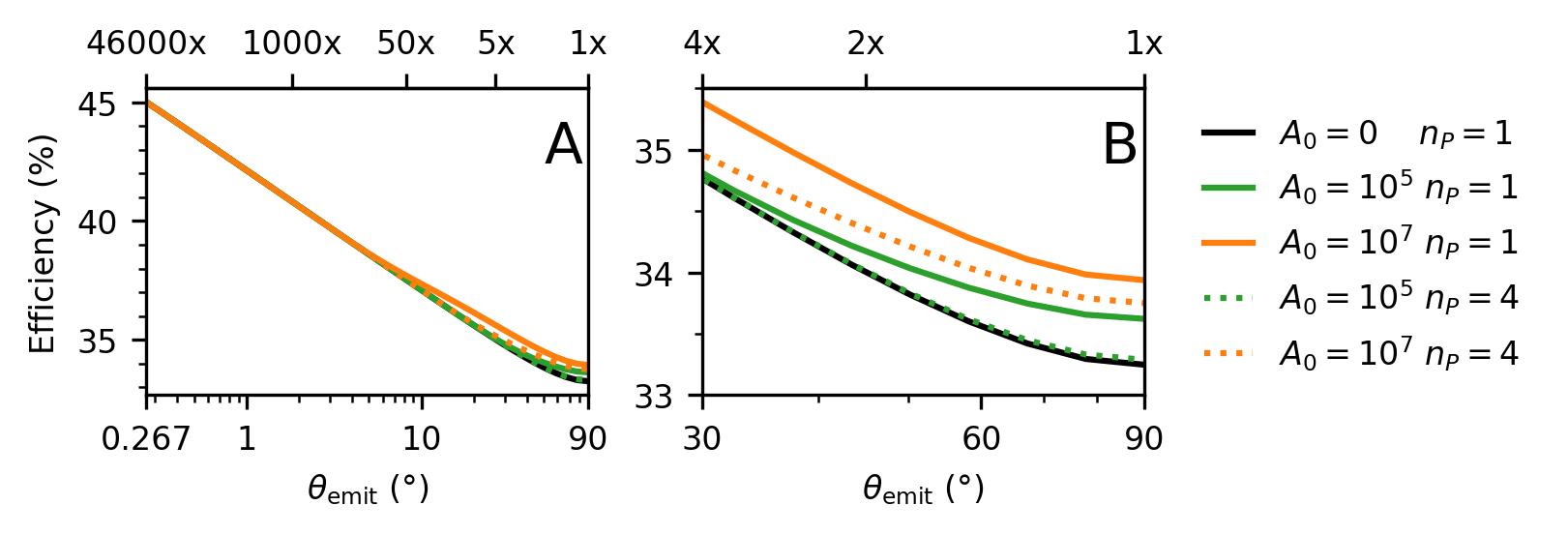}
    \caption{LAR performance as a function of angular restriction of emission. Conversion efficiency plotted versus the limiting emission half-angle, $\theta_{\mathrm{emit}}$, which restricts the solid angle of emission to a cone spanning polar angles $0 \leq \theta \leq \theta_{\mathrm{emit}}$ measured from the surface normal. Calculations are performed for absorption coefficients of the LAR medium, $A_0 = 10^5\,\mathrm{m}^{-1}$ (green) and $A_0 = 10^7\,\mathrm{m}^{-1}$ (orange), and for PV refractive index $n_P = 1$ (solid) and $n_P = 4$ (dotted). In all cases, the refractive index of the LAR medium is fixed at $n_L = 1$. The flat-plate detailed-balance reference is provided (black). Panel (A) spans emission restrictions from $\theta_{\mathrm{emit}} = 90^\circ$ (unrestricted hemispherical emission; 1x solar concentration factor) to $\theta_{\mathrm{emit}} = 0.267^\circ$ (46,000x solar concentration factor). Panel (B) expands the regime of weak angular restriction, from $\theta_{\mathrm{emit}} = 90^\circ$ (1x solar concentration factor) to $\theta_{\mathrm{emit}} = 30^\circ$ (4x solar concentration factor).}
    \label{fig:tAR}
\end{figure}

Figure~\ref{fig:tAR} examines the dependence of conversion efficiency on the imposed emission-cone half-angle, with LAR refractive index $n_L=1$, $A_0=10^5$ or $10^7~\mathrm{m}^{-1}$, and PV refractive index $n_P=1$ or $4$. Panel A spans unrestricted hemispherical emission at $90^\circ$ to the ideal concentration endpoint at $0.267^\circ$, corresponding to approximately $46{,}000\times$ equivalent solar concentration. Narrowing the emission cone raises the total efficiency of both the LAR configurations and the inactive flat-plate reference. The curves converge near the strongest angular restriction, where the additional LAR benefit becomes negligible. Panel B expands the weak-restriction regime, $30^\circ$--$90^\circ$, revealing the larger enhancement provided by the higher-loading LAR relative to the flat-plate reference at the same emission angle.

A noteworthy feature is the dependence on the PV refractive index, which does not enter the conventional flat-plate detailed-balance benchmark explicitly when the external absorptivity is prescribed. For the configurations shown, increasing $n_P$ from $1$ to $4$ reduces the additional LAR benefit, with the low-loading, high-index configuration remaining close to the reference. With $n_L=1$ fixed, increasing $n_P$ narrows the internal escape cone for PV luminescence entering the LAR. This changes the balance between photon recycling within the PV, radiative exchange with the LAR, and eventual escape into the environment. The PV refractive index therefore becomes a relevant design parameter for the coupled architecture even under ideal radiative conditions.

We next examine how material losses and imperfect nanorod alignment affect the predicted LAR benefit. We first consider non-unity nanorod QY (Figure 6) and imperfect alignment, using an azimuthally averaged tilt angle to approximate orientational disorder (Figure 7). We then evaluate these imperfections individually and together in a device that also includes Auger recombination in the PV (Figure 8), assessing how these practical constraints affect the efficiency gain relative to the corresponding flat-plate reference. See SI Sections 3-5 for the detailed mathematical procedure used to account for these sources of loss.

\begin{figure}
    \centering
    \includegraphics[width=1\linewidth]{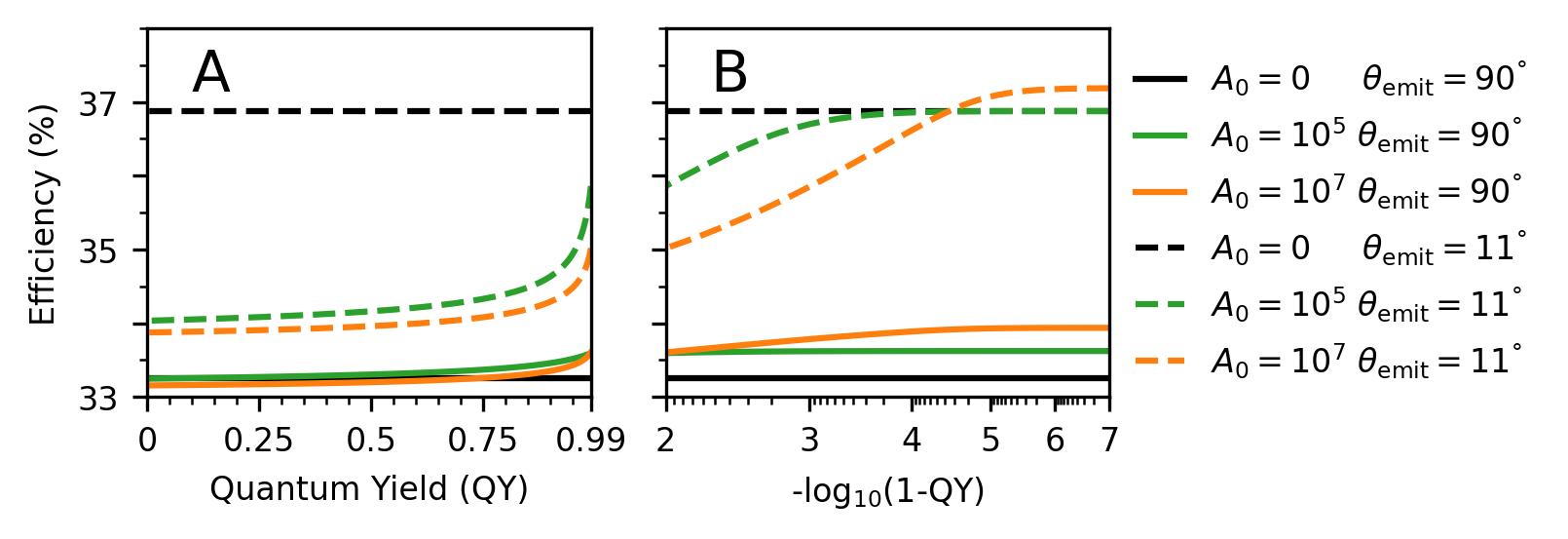}
    \caption{LAR performance as a function of nanorod photoluminescence quantum yield (QY). Conversion efficiency plotted versus QY for absorption coefficients of the LAR medium, $A_0 = 10^5\,\mathrm{m}^{-1}$ (green) and $A_0 = 10^7\,\mathrm{m}^{-1}$ (orange). The solid angle of emission is restricted to a cone spanning polar angles $0 \leq \theta \leq \theta_{\mathrm{emit}}$ measured from the surface normal. Solid and dashed traces correspond to limiting emission half-angles $\theta_{\mathrm{emit}} = 90^\circ$ (1x solar concentration factor) and $\theta_{\mathrm{emit}} = 11^\circ$ (approximately 27x solar concentration factor), respectively. The flat-plate detailed-balance references for each solar concentration factor are provided (black). Panel (A) plots efficiency for nanorod QY spanning 0--0.99, whereas panel (B) expands the near-unity QY regime.}
    \label{fig:plqy}
\end{figure}

Figure~\ref{fig:plqy} shows that nanorod QY strongly determines which loading regime is favorable. Panel (A) covers QYs from 0 to 0.99, while panel (B) resolves the near-unity regime using $-\log_{10}(1-\mathrm{QY})$, for which axis values of 2, 3, and 4 correspond to QYs of 99\%, 99.9\%, and 99.99\%, respectively. Solid and dashed traces correspond to emission half-angles of $90^\circ$ and $11^\circ$, with matching black flat-plate references. When QY is reduced, photons absorbed in the LAR have a finite probability of being lost nonradiatively rather than re-emitted back into useful radiative pathways. This penalty is most severe for high-loading layers, where larger $A_0$ increases both photon recycling and the number of absorption events in the nanorod medium. As a result, the low-loading configuration, $A_0=1\times10^{5}~\mathrm{m}^{-1}$, is more tolerant of reduced QY, while the high-loading configuration, $A_0=1\times10^{7}~\mathrm{m}^{-1}$, provides the larger efficiency gain only when QY remains near unity. Greater tolerance does not necessarily imply an efficiency gain: at $\theta_{\mathrm{emit}}=11^\circ$, both LAR configurations remain below the correspondingly restricted flat-plate reference throughout panel (A). In panel (B), the high-loading configuration exceeds this reference only very close to unity QY, while the low-loading configuration approaches it.

The required material performance is demanding. CdSe/CdS nanorods with near-unity QYs have been demonstrated,\supercite{CdSe} and photothermal-threshold quantum-yield measurements have resolved external luminescent efficiencies above 99.5\% in related CdSe/CdS quantum-dot systems.\supercite{Hanifi2019} These results motivate the pursuit of high-QY LAR materials, although the strongly angle-restricted cases impose substantially more stringent requirements.

\begin{figure}
    \centering
    \includegraphics[width=1\linewidth]{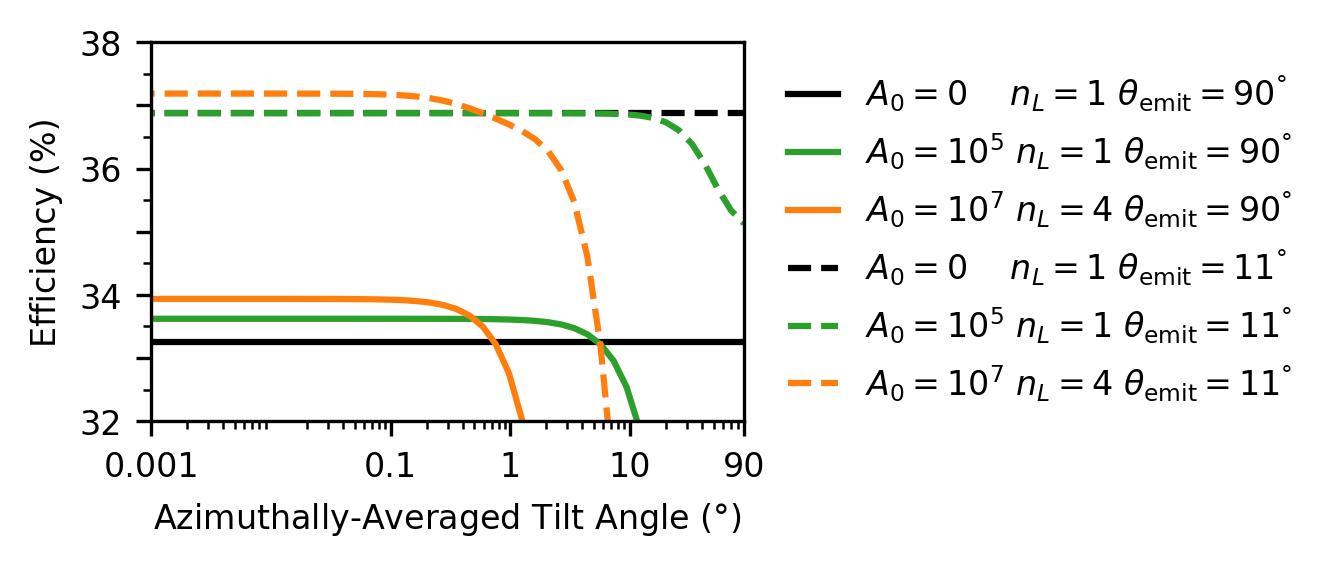}
    \caption{Sensitivity of LAR performance to uniform nanorod tilt. Conversion efficiency plotted versus the nanorod tilt angle, $\theta_{\mathrm{tilt}}$. The solid angle of emission is restricted to a cone with limiting emission half-angle $\theta_{\mathrm{emit}} = 90^\circ$ (solid; 1x solar concentration factor) or $\theta_{\mathrm{emit}} = 11^\circ$ (dashed; approximately 27x solar concentration factor). Calculations are performed for absorption coefficients of the LAR medium, $A_0 = 10^5\,\mathrm{m}^{-1}$ (green) and $A_0 = 10^7\,\mathrm{m}^{-1}$ (orange). The flat-plate detailed-balance references for each solar concentration factor are provided (black).}
    \label{fig:nanorod-tilt}
\end{figure}

Figure~\ref{fig:nanorod-tilt} examines the sensitivity of the LAR to orientational disorder in the nanorod ensemble. To approximate imperfect alignment, we assign the nanorods a common polar tilt angle $\theta_{\text{tilt}}$ relative to the surface normal and apply azimuthal averaging, while keeping the incident sunlight normal to the device. This parameter describes nanorod alignment rather than solar tracking error. The azimuthally averaged dipolar angular weighting function becomes
\begin{equation*}
f(\theta) = 1 - \left(\cos^2(\theta_{\text{tilt}})\cos^2(\theta) + \frac{1}{2}\sin^2(\theta_{\text{tilt}})\sin^2(\theta)\right).
\end{equation*}

The green curves represent $A_0=10^5~\mathrm{m}^{-1}$ and $n_L=1$, while the orange curves represent $A_0=10^7~\mathrm{m}^{-1}$ and $n_L=4$; solid and dashed traces correspond to emission half-angles of $90^\circ$ and $11^\circ$, respectively. Each configuration is compared with the flat-plate reference at the same angular restriction. The high-loading, high-index configuration provides the larger efficiency enhancement near perfect alignment, but this benefit rapidly diminishes as tilt increases. The low-loading, low-index configuration is more tolerant of tilt and retains a smaller enhancement over a broader tilt range under unrestricted emission. At $\theta_{\mathrm{emit}}=11^\circ$, however, the low-loading configuration remains closeer to the restricted flat-plate reference before declining.

Nanorod tilt directly compromises the angular selectivity that enables the LAR mechanism. Vertically aligned nanorods are nearly transparent to normally incident sunlight while still absorbing oblique PV luminescence. Tilting the ensemble increases absorption of near-normal sunlight in the LAR before it reaches the PV. Such absorption is not necessarily a terminal loss, because re-emission can return photons to the PV. Tilt also changes the angular distribution of re-emission and the balance between photon recycling and escape, weakening the efficiency enhancement. The results therefore identify a tradeoff between the ideal LAR benefit and tolerance to imperfect alignment. Electric-field alignment of colloidal semiconductor nanorods has been demonstrated,\supercite{RiviAlign,Mohammadimasoudi} but achieving the orientational control required for the largest gains remains a materials challenge.

\begin{figure}
    \centering
    \includegraphics[width=1\linewidth]{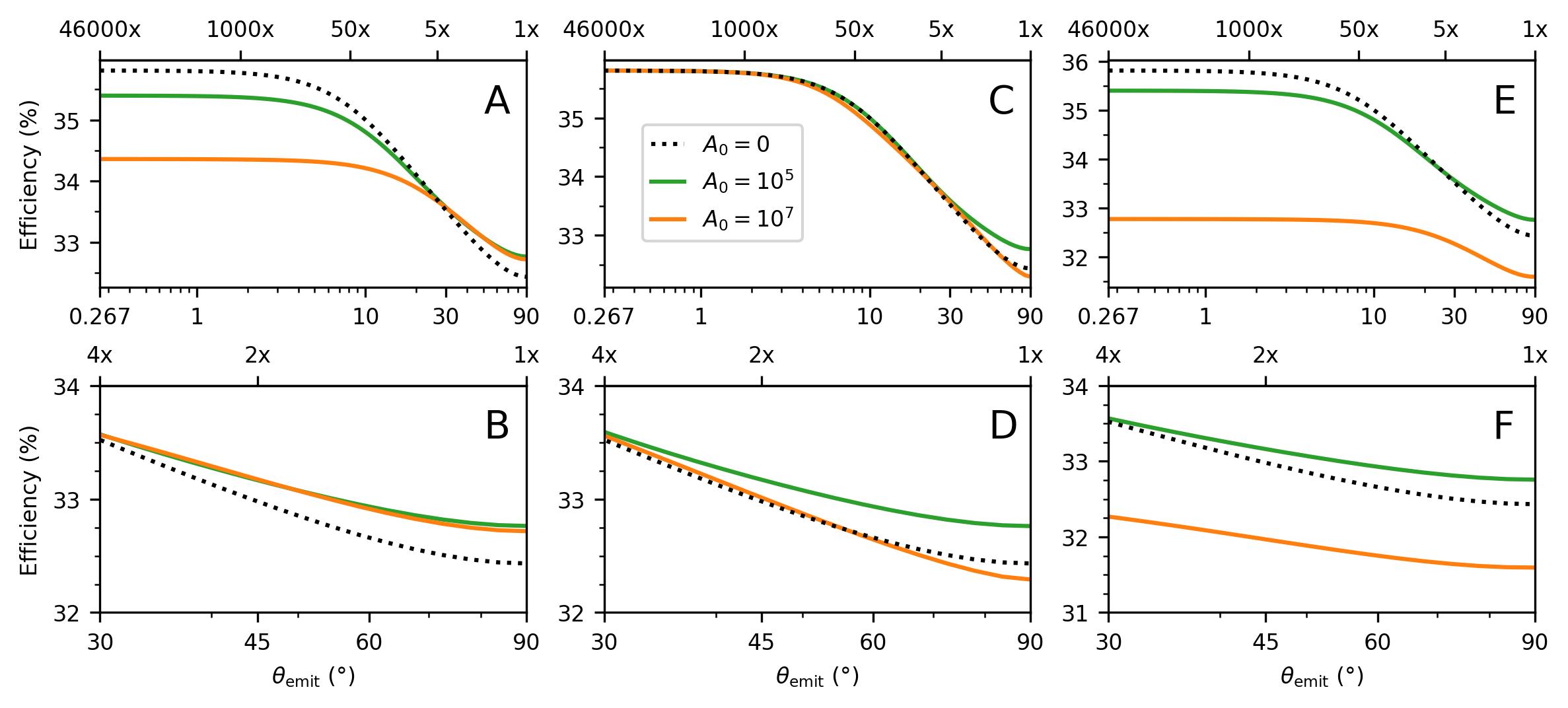}
    \caption{LAR performance as a function of angular restriction of emission with Auger recombination included as a non-radiative loss mechanism. Conversion efficiency plotted versus the limiting emission half-angle, $\theta_{\mathrm{emit}}$, which restricts the solid angle of emission to a cone spanning polar angles $0 \leq \theta \leq \theta_{\mathrm{emit}}$ measured from the surface normal. Panels (A,B) consider non-unity nanorod photoluminescence quantum yield ($\mathrm{QY} = 0.999$). Panels (C,D) consider non-ideal nanorod tilt ($\theta_{\mathrm{tilt}} = 0.5^\circ$). Panels (E,F) consider both non-unity QY and nanorod tilt ($\mathrm{QY} = 0.999$ and $\theta_{\mathrm{tilt}} = 0.5^\circ$). In each panel, calculations are performed for absorption coefficients of the LAR medium, $A_0 = 10^5\,\mathrm{m}^{-1}$ (green) and $A_0 = 10^7\,\mathrm{m}^{-1}$ (orange). In all calculations, the PV and LAR refractive indices are fixed at $n_P=n_L=1$. The flat-plate reference with $A_0 = 0\,\mathrm{m}^{-1}$ is provided (black, dotted).}
    \label{fig:Auger}
\end{figure}

After evaluating nonidealities associated with the LAR layer, we next include Auger recombination as an intrinsic PV-side nonradiative pathway. In an Auger process, the energy released by electron-hole recombination is transferred to a third carrier rather than emitted as a photon, reducing the radiative flux available for photon recycling.\supercite{WeisbergAuger,WeisbergAugerErratum,Strauss1993} Figure~\ref{fig:Auger} considers a $1~\mu\mathrm{m}$ GaAs PV absorber with $E_g=1.42~\mathrm{eV}$ and $n_P=n_L=1$. All cases include Auger recombination. Panels (A,B) examine $\mathrm{QY}=0.999$, panels (C,D) examine $\theta_{\mathrm{tilt}}=0.5^\circ$, and panels (E,F) include both imperfections. The upper panels span $\theta_{\mathrm{emit}}=0.267^\circ$--$90^\circ$, while the lower panels expand the weak-restriction range of $30^\circ$--$90^\circ$. The LAR configurations, $A_0=10^5~\mathrm{m}^{-1}$ (green) and $10^7~\mathrm{m}^{-1}$ (orange), are compared with the corresponding flat-plate reference (black, dotted).

The combined-loss cases refine the design tradeoffs identified in the separate sensitivity analyses. With non-unity QY alone, both LAR configurations exceed the reference under weak angular restriction, but fall below it under strong restriction, with a larger penalty at high loading. With nanorod tilt, the low-loading configuration retains a modest benefit under weak restriction, whereas the high-loading configuration falls below the reference toward unrestricted emission. When both imperfections are included, the low-loading LAR still provides a benefit under weak restriction, while the high-loading LAR remains below the reference throughout the plotted range. Lower loading therefore provides greater tolerance to imperfections, at the cost of a smaller ideal LAR enhancement. Realizing the larger ideal gains associated with high loading requires QY closer to unity and tighter nanorod alignment.

Together, Figures~\ref{fig:tAR} and~\ref{fig:Auger} also connect the LAR benefit to the angular restriction required for a given performance target. Where the LAR improves efficiency, a given normal-incidence conversion efficiency can be reached with a wider permitted emission cone than the corresponding flat-plate reference. This suggests a possible relaxation of the tracking requirements imposed by conventional angle-restricting or concentrating optics. The comparison does not directly establish tracking-error tolerance, however, because the intrinsic response of the LAR to off-axis illumination must also be considered.

All reported efficiencies use AM1.5D direct illumination and are normalized to the incident power in that direct component. Diffuse-sky radiation contributes neither to photogeneration nor to the incident-power normalization. A wider angular allowance could admit more off-axis radiation, but useful collection depends on nanorod absorption, re-emission, nonradiative loss, and escape. Restricting luminescent emission therefore involves a tradeoff with collecting diffuse light, as in conventional angle-restricting and concentrating systems. Establishing the net benefit under combined direct and diffuse illumination requires a specified spectral and angular sky distribution and lies beyond the present analysis.

The $1~\mu\mathrm{m}$ GaAs example provides a thin-absorber geometry for examining these competing optical and recombination processes. Auger loss is evaluated using the carrier density associated with the operating voltage, while electrical extraction remains simplified. The model does not resolve junction electrostatics, spatial carrier transport, or surface and contact losses. Choosing a thin absorber does not itself establish ideal carrier collection. The resulting efficiencies should therefore be interpreted as conditional performance estimates under the stated optical and electrical assumptions, rather than predictions for a fully specified device.

Taken together, these calculations predict that an LAR efficiency advantage can persist when Auger recombination, non-unity nanorod QY, and imperfect alignment are included, provided that nanorod loading and angular restriction are appropriately chosen. These results support a plausible route toward practical photovoltaic efficiency improvements and identify material and optical design conditions that favor realizing that benefit.

\section{Conclusions}

We used a Markov-chain radiative-transport model to evaluate a luminescent angle restrictor coupled to a conventional flat-plate photovoltaic cell. The model preserves microscopic reversibility for local optical transitions while allowing the device-level emission profile to emerge from coupled photon transport. The calculation resolves spatially distributed, voltage-dependent local emission rates and combines these contributions with fixed angular escape probabilities to determine the external emission profile, consistent with linear superposition.

Under idealized conditions, the LAR increases conversion efficiency by narrowing the angular distribution of escaping luminescence while maintaining solar absorption near normal incidence. The additional benefit over a flat-plate reference at the same emission-cone half-angle is greatest at lower band gaps and with weaker angular restriction. The calculations identify nanorod concentration and interfacial refractive-index contrast as central design parameters that jointly govern photon exchange, recycling, and escape. Including nonidealities reveals a tradeoff: lower nanorod loading offers greater tolerance to non-unity QY and imperfect alignment but a smaller ideal efficiency gain, whereas approaching the largest ideal gains requires higher QY and stronger orientational order. 

These results do not replace the detailed-balance framework. Instead, they identify a coupled radiative geometry in which the operating emission profile must be calculated from the coupled local emission rates rather than prescribed from equilibrium absorptivity alone. The LAR therefore provides both a proposed route for controlling luminescent emission in photovoltaics and a test case for applying Markov-chain radiative-transport methods to photovoltaic systems with more complex internal light exchange.

\section{Methods}

\subsection{Radiative-balance framework}

At thermal equilibrium, a photovoltaic (PV) cell in an isothermal blackbody environment obeys Kirchhoff's law: its spectral and angular emissivity equals its absorptivity, and the absorbed and emitted ambient photon fluxes balance. Retaining the angular dependence explicitly, the ambient blackbody photon flux is
\begin{equation}
    N_{\mathrm{amb}}
    =\frac{2}{h^3c^2}
      \int_{E_g}^{\infty}\!dE\int_0^{2\pi}\!d\phi
      \int_0^{\theta_{\mathrm{emit}}}\!d\theta\,
      a(\theta,\phi,E)
      \frac{E^2}{\exp(E/k_BT)-1}
      \cos\theta\sin\theta,
    \label{eq:ambient-flux}
\end{equation}

where $a(\theta,\phi,E)$ is the absorptivity, $E$ is photon energy, and $T$ is the cell temperature. Here, \(\theta_{\mathrm{emit}}\) defines the maximum polar half-angle of the solid angle into which the body emits radiation and, equivalently, the maximum polar half-angle of the solid angle over which the body absorbs ambient radiation.  At equilibrium, the emitted and absorbed ambient fluxes are equal and $e(\theta,\phi,E)=a(\theta,\phi,E)$.\supercite{Kats} 

For a PV reference emitter with a spatially uniform electron--hole quasi-Fermi-level splitting, the photon chemical potential is $\mu_\gamma=qV$, where $V$ is the PV operating voltage. The generalized Planck expression for its external emitted photon flux per unit area is
\begin{equation}
    N_{\mathrm{emit}}(V)
    =\frac{2}{h^3c^2}
      \int_{E_g}^{\infty}\!dE\int_0^{2\pi}\!d\phi
      \int_0^{\theta_{\mathrm{emit}}}\!d\theta\,
      e(\theta,\phi,E)
      \frac{E^2}{\exp[(E-qV)/(k_BT)]-1}
      \cos\theta\sin\theta,
    \label{eq:generalized-planck}
\end{equation}
where $e(\theta,\phi,E)$ is the spectral and angular emissivity of this PV reference emitter and $q$ is the elementary charge.\supercite{Wurfel1}

In the coupled architecture of the PV and luminescent angle restrictor (LAR), a common photon chemical potential $qV$ is not assumed throughout the composite. Local steady-state emission rates in the LAR are determined by absorption and coupled radiative exchange at each operating point. For fixed optical parameters, the radiative transport obeys linear superposition. External angular emission is obtained by summing the local radiative contributions with their depth-dependent angular escape probabilities, without assuming uniform excitation throughout the LAR. These calculations are described in the SI section 2. The absorbing-chain Markov construction below determines the photogeneration and recombination-loss terms entering the current--voltage relation.

\subsection{Polarization-resolved optical coupling}

Unpolarized illumination can be represented as an equal-intensity combination of orthogonal \(s\)- and \(p\)-polarized channels. Although the PV is modeled as an optically isotropic medium, the luminescent angle restrictor (LAR) exhibits an anisotropic optical response arising from the orientation-dependent coupling between the electric field and the transition dipole moments of its constituent nanorods (NRs). Consequently, the polarization dependence of the optical interactions must be considered explicitly.

The refractive indices of the PV, LAR medium, and external environment are denoted by $n_P$, $n_L$, and $n_E$, respectively.

For the ideal aligned-rod geometry, each nanorod is oriented along the surface normal, taken to be $\hat{z}$. Its electric dipole moment is therefore $\mathbf{p}=p_z\hat{z}$. Using complex phasors with time dependence $e^{-i\omega t}$, the time-averaged power absorbed by an electric dipole is proportional to
\begin{equation}
    P_{\abs}=\frac{\omega}{2}\operatorname{Im}\!\left(\mathbf{p}\cdot\mathbf{E}^{*}\right),
    \label{eq:dipole-absorbed-power}
\end{equation}
where $\mathbf{E}$ is the local electric field. The induced dipole is related to the local field through the polarizability tensor, $\mathbf{p}=\boldsymbol{\alpha}\mathbf{E}$. For a dipole constrained to the $z$ direction, the relevant polarizability is $\boldsymbol{\alpha}=\alpha_{zz}\hat{z}\hat{z}$, so that
\begin{align}
    \mathbf{p}&=\alpha_{zz}(\hat{z}\cdot\mathbf{E})\hat{z},\\
    P_{\abs}&\propto\operatorname{Im}(\alpha_{zz})\left|E_z\right|^2.
    \label{eq:dipole-z-coupling}
\end{align}
Thus, an ideally aligned nanorod couples only to the component of the optical electric field parallel to its axis.

For a plane wave propagating in the $xz$-plane at zenith angle $\theta$ relative to $+z$, the propagation direction and the corresponding polarization vectors may be written as
\begin{align}
    \hat{\mathbf{k}}&=\sin\theta\,\hat{x}+\cos\theta\,\hat{z},\\
    \mathbf{E}_{s}&=E_{0,s}\hat{y},\\
    \mathbf{E}_{p}&=E_{0,p}\left(\cos\theta\,\hat{x}-\sin\theta\,\hat{z}\right).
    \label{eq:sp-fields}
\end{align}
The $s$-polarized field is perpendicular to the plane of incidence and has no $z$ component. The $p$-polarized field lies in the plane of incidence and has $E_{p,z}=-E_{0,p}\sin\theta$. Substitution into Eq.~\eqref{eq:dipole-z-coupling} therefore gives
\begin{align}
    P_{\abs,s}&=0,\\
    P_{\abs,p}&\propto |E_{0,p}|^2\sin^2\theta.
    \label{eq:aligned-rod-polarization}
\end{align}
The angular factor $\sin^2\theta$ uses the propagation angle inside the LAR. For solar illumination centered on the surface normal, the solar disk subtends an external half-angle of $0.267^{\circ}$; the corresponding internal angles follow Snell's law, Eq.~\eqref{eq:snell-law}. Near-normal propagation suppresses the angular coupling factor, while absorption and transmission also depend on $A_0$, layer thickness, and the interface and propagation probabilities.

If the nanorods are tilted by an angle $\beta$ relative to $\hat{z}$, their dipole axes acquire in-plane components. For an ensemble with a common tilt angle and azimuths uniformly distributed in the plane, azimuthal averaging gives the polarization-dependent coupling factors
\begin{align}
    f_{L,p}(\theta,\beta)
    &=\frac{1}{2}\sin^2\!\beta\cos^2\!\theta
      +\cos^2\!\beta\sin^2\!\theta,
      \label{eq:tilted-p-coupling}\\
    f_{L,s}(\theta,\beta)
    &=\frac{1}{2}\sin^2\!\beta.
    \label{eq:tilted-s-coupling}
\end{align}
The corresponding NR absorption coefficients are $\alpha_{L,p}(\theta)=A_0 f_{L,p}(\theta,\beta)$ and $\alpha_{L,s}(\theta)=A_0 f_{L,s}(\theta,\beta)$, where $A_0$ is the nanorod absorption-coefficient scale, with units of $\mathrm{m}^{-1}$. For a fixed nanorod absorption cross-section and material model, increasing $A_0$ corresponds to increasing nanorod loading. The aligned-rod limit is recovered at $\beta=0$, for which $f_{L,p}=\sin^2\theta$ and $f_{L,s}=0$. For ideally vertical nanorods, increasing $A_0$ does not introduce absorption of the $s$-polarized component. Increasing nanorod loading therefore does not eliminate this uncoupled polarization channel.

The same orientation projections determine the polarization-resolved dipole-emission weights by reciprocity. The corresponding dipole angular dependence is consistent with angular-emission treatments for luminescent nanostructures.\supercite{Alivisatos_lin_pol,RiviDB,RodriguezOrtiz2023} For the PV, which is treated as an isotropic emitter, the angular weighting is taken to be unity in each normalized polarization calculation:
\begin{equation}
    f_{P,p}(\theta)=f_{P,s}(\theta)=1
    \label{eq:PV-coupling}
\end{equation}
The optical response of the LAR depends on the polarization of the incident optical field. Accordingly, the s- and p-polarized contributions to absorption and propagation are evaluated separately. However, unlike a polarization-preserving luminescent medium, the nanorods do not retain the polarization of the absorbed photon upon re-emission. Instead, each radiative re-emission event produces s- and p-polarized components according to the jointly normalized dipole emission distribution. Consequently, the two polarization contributions are combined to construct a single transition matrix, \(\mathbf P_{\mathrm{mix}}\). For LAR-originating emission, the s- and p-polarized contributions are summed, whereas environment- and PV-originating source columns are averaged to represent their equal s/p populations.

\subsection{Optical transition matrix}

The coupled architecture is discretized along the surface-normal direction. The external environment is represented by state $1$, the LAR by states $2$ through $N_L+1$, and the PV by states $N_L+2$ through $N_{\mathrm{tot}}$, where
\begin{equation}
    N_{\mathrm{tot}}=1+N_L+N_P.
    \label{eq:number-of-states}
\end{equation}
Each material region is divided into layers of thickness $dz$. We denote the LAR and PV state sets by $\mathcal{L}=\{2,\ldots,N_L+1\}$ and $\mathcal{P}=\{N_L+2,\ldots,N_{\mathrm{tot}}\}$, respectively.

For each polarization \(\pol\), we construct a polarization-resolved contribution \(\mathbf P_\pol\) to the optical transition probabilities. These contributions are subsequently assembled into the combined transition matrix \(\mathbf P_{\mathrm{mix}}\), according to the source-dependent polarization normalization described above. Further details of the emission, absorption, and transmission functions are provided in the SI section 1.

The optical transition matrix is indexed by destination-state rows and source-state columns. For each physically allowed source state, the probabilities of all subsequent optical outcomes, including absorption in another layer and escape to the environment, are normalized to unity. 
The s- and p-polarized components of LAR-originating emission are treated as contributions to a jointly normalized dipole emission distribution. Accordingly, these contributions are summed when constructing the LSC source columns of the combined transition matrix, while the environment- and PV-originating columns are averaged to account for unpolarized illumination. The resulting combined transition matrix is column-stochastic. In the jointly normalized case
\begin{equation}
    \sum_j P_{ji,s}+\sum_j P_{ji,p}=1 \qquad \text{for } i \in\mathcal{L}_{\actst}
    \label{eq:joint-norm-column-stochastic}
\end{equation}

Likewise, for environment- and PV-originating states,
\begin{equation}
    \frac{1}{2}\sum_j P_{ji,s}+\frac{1}{2}\sum_j P_{ji,p}=1, \qquad \text{for } i\in\{1\}\cup\mathcal{P}
\end{equation}

The polarization-resolved optical contributions are assembled into a single transition matrix, \(\mathbf P_{\mathrm{mix}}\), according to the normalization appropriate to each source state:
\begin{equation}
    \Pmix(:,i)=\begin{cases}
        \mathbf{P}_s(:,i)+\mathbf{P}_p(:,i), \qquad i\in\mathcal{L}_{\text{active}}\\
        \frac{1}{2}[\mathbf{P}_s(:,i)+\mathbf{P}_{p}(:,i)],\qquad i\in\{1\}\cup\mathcal{P}
    \end{cases}
    \label{eq:P_mix_form}
\end{equation}

The polarization-resolved transition probabilities are constructed by integrating over the allowed emission angles and optical propagation paths,
\begin{equation}
    P_{ji,\pol}
    =\int_0^{\pi/2}
      \mathrm{d}p^{(e)}_{i,\pol}(\theta)
      \,p^{(t)}_{\pol}(\theta,i,j)
      \,p^{(a)}_{j,\pol}(\theta),
    \label{eq:transition-probability}
\end{equation}
where \(dp^{(e)}_{i,\pol}(\theta)\) is the differential angular emission weight associated with source state \(i\) and polarization \(\pol\), \(p^{(t)}_{\pol}(\theta,i,j)\) accounts for the allowed optical propagation paths from source state \(i\) to destination state \(j\), and \(p^{(a)}_{j,\pol}(\theta)\) is the absorption factor associated with the destination state. The propagation contributions include direct transmission, reflection from the back mirror, and paths involving total internal reflection, as applicable. When multiple optical paths connect a given pair of states, their contributions are summed in constructing the corresponding transition probability. The angular emission weights are normalized over the full sphere, whereas the transition-probability integrals are evaluated over a single hemisphere, with the distinct upward- and downward-emission trajectories accounted for through the optical path contributions. For LAR-originating emission, the s- and p-polarized angular weights are jointly normalized, rather than normalized independently for each polarization. For the environmental source state, the angular distribution is instead defined by the incident solar illumination, with its angular range restricted to the solar acceptance cone.

After integrating over azimuth, the differential angular emission weights for a specified source state and polarization are given by
\begin{equation}
    dp^{(e)}_{i,\pol}(\theta)
    =2\pi \nu f_{i,\pol}(\theta)\sin\theta\,d\theta
    \label{eq:angular-emission-probability}
\end{equation}
where $\nu=\nu^{(L)}=\frac{3}{8\pi}$ when the emission occurs within the NR medium, and $\nu=\nu^{(P)}=\frac{1}{4\pi}$ when emission occurs within the PV. For PV emission, $f_{i,\pol}(\theta)=1$, as given by Eq.~\eqref{eq:PV-coupling}. For NR emission within the LAR, the polarization-dependent dipole weights are given by Eqs.~\eqref{eq:tilted-p-coupling} and~\eqref{eq:tilted-s-coupling}.
The normalization procedure for the PV and the NR medium differs in that the angular emission distribution for each polarization the PV is normalized independently according to
\begin{equation}
    \int_0^\pi 2\pi \nu^{(P)}\sin(\theta)\mathrm{d}\theta =1
    \label{eq:PV-normalization}
\end{equation}
whereas the angular emission distribution for the NR medium is normalized according to
\begin{equation}
    \sum_{\pol\in\{s,p\}}\int_0^\pi 2\pi \nu^{(L)}f_{i,\pol}(\theta)\sin(\theta)\mathrm{d}\theta=1
\end{equation}

If the angular emission weight for an individual polarization component vanishes, that component contributes no emission to the corresponding polarization channel. This does not necessarily render the LAR state inactive, since the other polarization component may retain a nonzero emission weight. We define a LAR state as physically active only when it possesses a nonzero absorption coupling and a nonzero jointly normalized dipole emission distribution. States for which the LAR absorption coefficient vanishes, such as the \(A_0=0\) limit, are therefore excluded from the physical active state set. In the numerical implementation, the full LAR state array is retained for consistent indexing. States with vanishing optical coupling may remain as bookkeeping placeholders, but they are not populated through physical optical transitions and do not contribute to the calculated photogeneration or recombination-loss terms. The corresponding first-passage probabilities are physically interpreted only for active LAR states.

Absorption and single-layer transmission are described by the Beer--Lambert law. For a layer of thickness $dz$, the transmission and absorption probabilities at propagation angle $\theta$ are
\begin{align}
    p^{(t)}_{\pol}(\theta)
    &=\exp\!\left[-\alpha_{\pol}(\theta)\frac{dz}{\cos\theta}\right],
    \label{eq:beer-lambert-transmission}\\
    p^{(a)}_{\pol}(\theta)
    &=1-\exp\!\left[-\alpha_{\pol}(\theta)\frac{dz}{\cos\theta}\right].
    \label{eq:beer-lambert-absorption}
\end{align}

The absorption probability is the complement of the transmission probability, $p^{(a)}_{\pol}(\theta)=1-p^{(t)}_{\pol}(\theta)$.

In the PV, absorption is isotropic, so $\alpha_{P,s}(\theta)=\alpha_{P,p}(\theta)=\alpha_P$. In the LAR, the polarization-dependent absorption coefficients are $\alpha_{L,\pol}(\theta)=A_0 f_{L,\pol}(\theta,\beta)$, where the dimensionless angular factors $f_{L,\pol}$ are given by Eqs.~\eqref{eq:tilted-p-coupling} and~\eqref{eq:tilted-s-coupling}. 
Self-absorption within an emitting layer must also be included because the device is discretized into finite layers. For emission from the center of a layer, the path to either face is $dz/(2\cos\theta)$. Self-absorption within the NR medium and the PV is defined as
\begin{equation}
\begin{aligned}
    p_{S,\pol}^{(a)}(\theta)
    &=2
      \left[1-\exp\!\left(-\frac{\alpha_{\text{medium},\pol}(\theta)dz}{2\cos\theta}\right)\right]
\end{aligned}
    \label{eq:self-absorption}
\end{equation}

Here, the factor of two accounts for the equivalent contributions from upward and downward emission trajectories originating at the center of a layer. Although the angular emission distribution is normalized over the full sphere, the transition probabilities are evaluated using angular integrals over a single hemisphere, with the optical path contributions accounting for both propagation directions. For self-absorption, the two directions have identical path lengths and absorption probabilities, allowing their contributions to be combined through the factor of two. Accordingly, \(p_{S,\mathrm{pol}}^{(a)}\) represents an effective self-absorption factor within the hemisphere-reduced integration, rather than the absorption probability of an individual ray. The absorption coefficient \(\alpha_{\mathrm{medium},\mathrm{pol}}\) is replaced by either \(\alpha_{L,\mathrm{pol}}\) or \(\alpha_{P,\mathrm{pol}}\), depending on whether self-absorption occurs within the NR medium or PV, respectively.

Propagation between layers includes direct transmission, transmission following reflection from the back mirror, and paths involving total internal reflection at either the PV--LAR or LAR--environment interface. The corresponding path expressions are given in SI section 1.4.

Refracted angles are determined by Snell's law,
\begin{equation}
    \theta_Y=\sin^{-1}\!\left(\frac{n_X}{n_Y}\sin\theta_X\right),
    \label{eq:snell-law}
\end{equation}
where $\theta_X$ is the incident angle in the initial medium, $\theta_Y$ is the refracted angle in the final medium, $n_X$ is the refractive index of the initial medium, and $n_Y$ is the refractive index of the final medium. If no propagating refracted solution exists, the corresponding path is treated as total internal reflection.

This construction retains the reciprocal optical rules for passive propagation and interfaces while allowing the NR absorption and emission response to depend on polarization.

\subsection{Absorbing Markov-chain formulation}

The combined optical transition matrix describes the pathways available to photons within the coupled PV--LAR--environment system. To determine the eventual fate of photons absorbed in the LAR, we employ an absorbing Markov-chain formulation, following established applications of Markov chains to radiative transport in luminescent and nanophotonic systems.\supercite{RiviMarkov} This construction determines the probabilities of eventual PV absorption, environmental escape, and nonradiative LAR quenching. Separately, the local steady-state LAR emission rates and their contributions to external angular emission are obtained through coupled radiative exchange and linear superposition, as described in the SI section 2.

For clarity, let $\Pmat_{\mix}(A,B)$ denote the submatrix whose destination-state rows belong to set $A$ and whose initial-state columns belong to set $B$. Let $Y_L$ be the photoluminescence quantum yield of the LAR.

Let \(\mathcal L_{\mathrm{active}}\subseteq\mathcal L\) denote the subset of LAR states that can be physically populated through optical absorption and possess a nonzero jointly normalized dipole emission distribution. Activity is determined by the combined optical response rather than by either polarization component individually. Consequently, an LAR state remains active when its coupling to one polarization vanishes, provided that coupling to the other polarization remains nonzero. For example, ideally vertical nanorods do not couple to the s-polarized electric field, but retain a nonzero p-polarized coupling, allowing the combined state to remain active. Conversely, states for which the LAR absorption coefficient vanishes, such as the \(A_0=0\) limit, do not couple to either polarization, and are therefore excluded from the physical active state set. The first-passage calculation begins with an absorbed excitation, before the subsequent radiative re-emission or nonradiative quenching decision.
The matrix of transitions that remain within the active LAR state set after radiative re-emission is
\begin{equation}
    Q=Y_L\Pmat_{\mix}(\mathcal{L}_{\text{active}},\mathcal{L}_{\text{active}}).
    \label{eq:Q-matrix}
\end{equation}

The probability vector for an excitation in an active LAR state to undergo radiative re-emission followed by environmental escape before another absorption is
\begin{equation}
    s_{\mathrm{esc}}
    =Y_L\Pmat_{\mix}(1,\mathcal{L}_{\text{active}})^{\mathsf T},
    \label{eq:s-escape}
\end{equation}
and the probability vector for nonradiative loss associated with a non-unity NR quantum yield is
\begin{equation}
    s_{\mathrm{q}}=(1-Y_L)\one_{L}.
    \label{eq:s-quench}
\end{equation}
Here, $\one_{L}$ is the vector of ones on $\mathcal{L}_{\text{active}}$, with $|\mathcal{L}_{\text{active}}|$ components. The identity matrix $I$ in the first-passage equations acts on the same active state set. The total one-step probability of leaving the LAR radiative subnetwork without first entering the PV is therefore
\begin{equation}
    s=s_{\mathrm{esc}}+s_{\mathrm{q}}.
    \label{eq:s-total}
\end{equation}

We define $\phi$ as the vector whose component for an active LAR state $i\in\mathcal{L}_{\text{active}}$ is the probability that an excitation following absorption in that state ultimately terminates by environmental escape or NR quenching before PV absorption. Because destination states index rows and source states index columns, the backward equation for a probability conditioned on its initial state involves the transpose of $Q$. The absorbing-chain equation is
\begin{equation}
    \left(I-Q^{\mathsf T}\right)\phi=s,
    \qquad\text{or equivalently}\qquad
    \phi=\left(I-Q^{\mathsf T}\right)^{-1}s.
    \label{eq:phi}
\end{equation}

The inverse has the Neumann-series representation
\begin{equation}
    (I-Q^{\mathsf T})^{-1}
    =\sum_{n=0}^{\infty}(Q^{\mathsf T})^{n}
    =I+Q^{\mathsf T}+(Q^{\mathsf T})^{2}+\cdots,
    \label{eq:neumann-series}
\end{equation}
which sums all possible numbers of repeated transitions within the LAR before termination. 

The Neumann series converges when the spectral radius satisfies \(\rho(Q)<1\). For a finite-state Markov chain, this condition holds when all states represented by \(Q\) are transient, meaning that no closed recurrent class exists entirely within the LAR subnetwork. A closed communicating class is a set of mutually accessible states from which no transition to a state outside the class is possible. Such a class would allow photons to remain indefinitely within the LAR subnetwork, preventing convergence of the Neumann series. In the present model, radiative transitions within the LAR are accompanied by pathways leading to PV absorption, environmental escape, or nonradiative quenching. Provided every active LAR state can eventually reach one of these terminating outcomes, the LAR subnetwork contains no closed recurrent class, ensuring that \(\rho(Q)<1\) and that the Neumann series converges.

The complementary vector
\begin{equation}
    \psi=\one_{L}-\phi
    \label{eq:psi}
\end{equation}
therefore gives, for each active LAR state, the probability that an excitation following absorption ultimately leads to PV absorption, irrespective of the number of intervening LAR absorption and re-emission events. Because $s$ includes both environmental escape and nonradiative NR quenching, $\phi$ is the probability of termination outside the PV-recycling pathway, not exclusively the probability of far-field escape.

For a photon emitted from a PV layer, the corresponding component of $\lambda$ gives the probability of loss through environmental escape or nonradiative quenching in the LAR before reabsorption by the PV. It combines direct PV-to-environment escape with PV-to-LAR transitions followed by escape or quenching. Mathematically, this is represented by
\begin{equation}
    \lambda =\mathbf{P}_{\text{mix}}(1,\mathcal{P})^T+\mathbf{P}_{\text{mix}}(\mathcal{L}_{\text{active}},\mathcal{P})^{T}\phi
    \label{eq:lambda}
\end{equation}
where $\mathcal{P}$ refers to the set of PV states.

\subsection{Photogeneration and radiative recombination}

Let $N_{\mathrm{sun}}$ be the incident solar photon flux above the PV band gap per unit device area, in $\mathrm{m}^{-2}\,\mathrm{s}^{-1}$. For an incident spectrum expressed as spectral irradiance $S(E)$,
\begin{equation}
    N_{\mathrm{sun}}
    =\int_{E_g}^{\infty}\frac{S(E)}{E}\,dE,
    \label{eq:solar-photon-flux}
\end{equation}
with the corresponding wavelength-domain integral used when the input spectrum is tabulated as $S(\lambda)$ and $E=hc/\lambda$. We use the AM1.5D solar spectrum to represent direct solar radiation in all solar-illumination calculations.\supercite{NREL} Diffuse-sky radiation is excluded from solar photogeneration.
Using the combined polarization-mixed transition matrix, we define the solar-driven PV generation rate per unit device area, $G$, in $\mathrm{m}^{-2}\,\mathrm{s}^{-1}$, as the sum of all $E\rightarrow P$ transitions plus all $E\rightarrow L$ transitions that ultimately reach the PV, scaled by the incident solar flux, $N_{\text{sun}}$:
\begin{equation}
    G= \Bigg(\Big(\sum \mathbf{P}_{\text{mix}}(\mathcal{P},\{1\})\Big)+\mathbf{P}_{\text{mix}}(\mathcal{L}_{\text{active}},\{1\})^T \psi
    \Bigg)N_{\text{sun}}
    \label{eq:generation-rate}
\end{equation}
The resulting photogenerated current density is then
\begin{equation}
    J_{\text{gen}}=qG
    \label{eq:generation-current-density}
\end{equation}

The zero-bias photon-emission source from the PV is obtained by integrating the Planck distribution over photon energies above the band gap. For a PV layer of thickness $dz$, the source term used in the model is
\begin{equation}
    r_{0,i}
    =\frac{8\pi n_P^2\alpha_P dz}{h^3c^2}(k_BT)^3
      \int_{x_g}^{\infty}\frac{x^2}{e^x-1}\,dx,
    \qquad
    x=\frac{E}{k_BT},\quad x_g=\frac{E_g}{k_BT},
    \label{eq:r0}
\end{equation}
where $n_P$ is the PV refractive index, $\alpha_P$ is its absorption coefficient, $T$ is the cell temperature, and $r_{0,i}$ is the zero-bias photon-emission rate from PV layer $i$ per unit device area, in $\mathrm{m}^{-2}\,\mathrm{s}^{-1}$. 

This term is derived by first considering the equilibrium photon number density per unit energy, which is the product of the photonic density of states of the PV, $g_P(E)=\frac{8\pi n_{P}^3 E^2}{h^3 c^3}$ with the average photonic mode occupation number, $\langle n \rangle=\frac{1}{\exp(\frac{E}{k_B T})-1}$. The absorption rate per photon is $\alpha_P \frac{c}{n_P}$, in $\mathrm{s}^{-1}$. Multiplying the equilibrium photon number density per unit energy by this rate gives the photon-absorption rate per unit volume per unit energy, equal at thermal equilibrium to the corresponding photon-emission rate, 
\begin{equation}
    r_{0,vol}(E) = \frac{8\pi n_{P}^2 \alpha_P E^2}{h^3 c^2} \frac{1}{\exp(\frac{E}{k_B T})-1}
    \label{eq:per-vol-per-en-em-rate}
\end{equation}
Integrating ~\eqref{eq:per-vol-per-en-em-rate} over energy and multiplying by the layer thickness $\mathrm{d}z$ yields the per-layer thermal emission rate per unit surface area shown in ~\eqref{eq:r0}. Each source term $r_{0,i}$ is then collected into a vector, $\mathbf{r}_{0}$.

The effective zero-bias recombination-loss photon flux per unit device area is obtained by weighting each PV-layer emission rate by the corresponding component of $\lambda$. Multiplication by the elementary charge gives the saturation current density:
\begin{equation}
J_{0}=q\mathbf{r}_{0}^T \lambda
    \label{eq:J0}
\end{equation}

\subsection{Current-voltage relation and efficiency}

In the Boltzmann limit of the radiative model the voltage-dependent recombination current density follows the Shockley diode relation.\supercite{Atwater}
\begin{equation}
    J_{\text{loss}}(V)
    =J_{0}\left[\exp\!\left(\frac{qV}{k_BT}\right)-1\right].
    \label{eq:rad-diode-current}
\end{equation}

In this limit, the net current density is the photogenerated current density minus the voltage-dependent recombination current density:
\begin{equation}
    J(V)=qG-J_{0}\left[\exp\!\left(\frac{qV}{k_BT}\right)-1\right]
    \label{eq:polarized-current}
\end{equation}

With a spatially uniform PV quasi-Fermi-level splitting \(qV\), the emission source associated with each PV layer scales by the same voltage-dependent factor $F(V)=\frac{I(V)}{I(0)}$ such that $r_{i}(V) = r_{0,i} F(V)$.

Here, $I(V)$ is defined as
\begin{equation}
    I(V)=\int_{E_g}^{\infty}\frac{E^2}{\exp(\frac{E-qV}{k_B T})-1}\mathrm{d}E
\end{equation}
Thus, ~\eqref{eq:polarized-current} may be replaced with
\begin{equation}
    J(V)=qG-J_0\left[F(V)-1\right]
    \label{eq:J_V_F_V}
\end{equation}

At short circuit, $V=0$ and $J_{\text{loss}}=0$, so $J_{\mathrm{SC}}=qG$ within this radiative model.
In the Boltzmann limit, $F(V)\simeq\exp(qV/k_BT)$, recovering the exponential form above in ~\eqref{eq:polarized-current}, and the analytic open-circuit voltage is obtained by setting the current density to zero:
\begin{equation}
    V_{\mathrm{OC}}
    =\frac{k_BT}{q}
      \ln\!\left(1+\frac{qG}{J_{0}}\right).
    \label{eq:VOC}
\end{equation}
With the full Planck integral, setting the current density to zero gives
\begin{equation}
    \frac{I(V_{\mathrm{OC}})}{I(0)}
    =1+\frac{qG}{J_{0}}.
    \label{eq:VOC-planck}
\end{equation}
The current density used for numerical power and efficiency calculations retains this full voltage dependence.

For calculations including Auger recombination in the PV, the current--voltage treatment is described in the SI section 5.

The output power density at an operating voltage $V$ is
\begin{equation}
    P(V)=J(V)V,
    \label{eq:power-density}
\end{equation}
and the maximum power density is $P_{\max}=\max_V P(V)$. The conversion efficiency is
\begin{equation}
    \eta=\frac{P_{\max}}{P_{\mathrm{in}}},
    \label{eq:efficiency}
\end{equation}
where $P_{\mathrm{in}}$ is the incident direct solar power density associated with the AM1.5D spectrum. The LAR and reference devices are evaluated using the same direct-illumination convention.

\section*{Acknowledgements}

Jonathan McTague acknowledges support from the National Science Foundation (CHE 2404128). Rivi Ratnaweera and Matthew Sheldon acknowledge support from the Office of Naval Research Science \& Technology (ONR N000142512272).

\section*{Supporting Information}

Supporting Information. Detailed derivations of the Markov-chain transition probabilities, implementation of nonideal loss pathways, and additional calculation details (PDF).

\section*{Data Availability}

The data underlying this study are available in the article and its Supporting Information. Code required to reproduce the calculations is available at https://github.com/jonmct123/MarkovChainLAR. Additional simulation input files and analysis details are available from the corresponding author upon reasonable request.

\section*{Conflict of Interest}

The authors declare no competing financial interest.

\printbibliography

\clearpage

\thispagestyle{empty}

\begin{center}
{\large\bfseries For Table of Contents Only\par}
\vspace{1em}

    \includegraphics[width=1\linewidth]
    {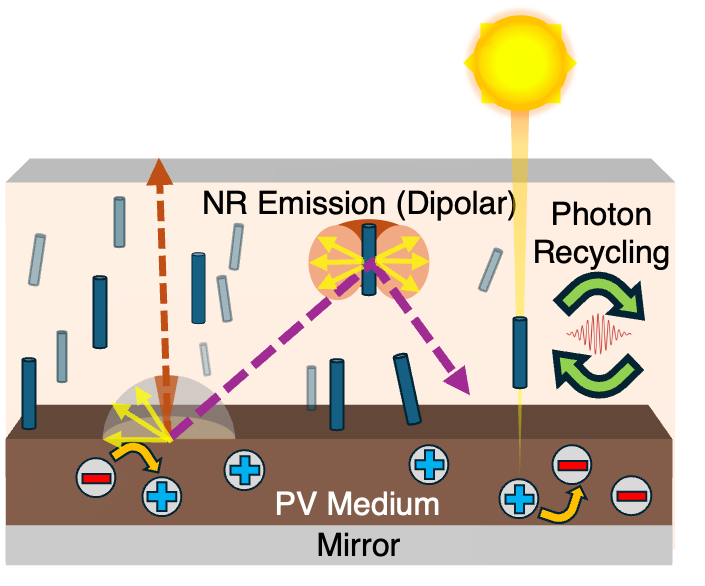}

\end{center}
\clearpage
\end{document}